\documentclass[twocolumn,superscriptaddress,showpacs]{revtex4}
\pdfoutput=1

\usepackage{graphicx,color}
\usepackage{amsmath}
\usepackage{dcolumn}

\def\mis{~$\rm{\mu ms^{-1}}$}
\def\degres{$^{\rm o}$}
\def\mic{$\rm{\mu{}m}$}  

\begin{document}

\newcommand{\INP}{
INSP, UPMC Univ. Paris 6, CNRS UMR 7588, 140 rue de Lourmel, 75015 Paris, France}

\newcommand{\SZFKI}{Research Institute for Solid State Physics and Optics,
                    POB 49, H-1525 Budapest, Hungary}

\title[Short Title]{Weakly faceted cellular patterns  {\it{versus}} growth-induced plastic deformation  in thin-sample directional solidification of monoclinic biphenyl}

\author{Tam\'as B\"orzs\"onyi}
\email{btamas@szfki.hu}
  \affiliation{\INP} 
  \affiliation{\SZFKI}

\author{Silv\`ere Akamatsu} 
  \affiliation{\INP} 

\author{Gabriel Faivre}
  \affiliation{\INP} 

\date{\today}

\begin{abstract}
We present an experimental study of  thin-sample directional solidification (T-DS) in  
impure biphenyl.  The plate-like growth shape of the monoclinic biphenyl 
crystals includes two low-mobility (001) facets and four high-mobility  \{110\} facets. Upon T-DS, biphenyl plates oriented with (001) facets parallel to the sample plane can  exhibit either a strong growth-induced plastic deformation (GID), or deformation-free weakly faceted  (WF) growth patterns. We determine the respective conditions of appearance of these phenomena. GID is shown to be a long-range thermal-stress effect, which disappears when the growth front has a cellular structure. An early triggering of the cellular instability allowed us to avoid GID and study the dynamics of WF patterns as a function of the orientation of the crystal.
\end{abstract}


\maketitle

\section{Introduction}

Directional solidification of dilute alloys gives rise to complex out-of-equilibrium growth patterns. The control of these patterns is a central issue in materials science \cite{Boettinger00} and raises fundamental problems in nonlinear physics. The basic phenomenon in the field is the bifurcation from a planar  to a digitate growth front, which occurs when the solidification rate $V$ exceeds a critical value $V_c\approx GD/\Delta T_o$, where $G$ is the applied thermal gradient, $D$ is the solute diffusion coefficient in the liquid and $\Delta T_o$ is the thermal gap of the alloy \cite{Rutter53, Mullins64}. The morphology of the fingers above the critical point  evolves   from rounded cells at  $(V-V_c)/V_c\ll1$  to  dendrites (parabolic tip and sidebranches) at  $(V-V_c)/V_c\gg1$  \cite{Kurowski90}. The dominant factors in the process are the diffusion of the chemical species in the liquid, and the resistance  of the solid-liquid interface to deformation, which is determined by $G$ and the physical properties of the interface itself, namely, its surface tension $\gamma$ and kinetic coefficient $\beta=\partial(\delta T_k)/\partial V$, where $\delta T_k$ is the kinetic undercooling.  While the value of $V_c$  is approximately  independent of $\gamma$ and $\beta$,  the characteristics of the cellular or dendritic patterns at $V>V_c$   crucially depend on these properties, especially, on their anisotropy  \cite{BenAmar86,Barbieri87,Brener92,Ihle94,Akamatsu95,Akamatsu97}.  A fundamental distinction must be made between nonfaceted  and faceted alloys, the latter being the alloys, in which $\gamma({\bf n})$ and/or $\beta({\bf n})$ have cusp singularities for some orientations of  {\bf n}, where  {\bf n} is the normal to the interface  referred to the crystalline axes.  The distinction between two-dimensional (or thin) and three-dimensional (or bulk) solidification is also important.  

\begin{figure}[htbp]
\includegraphics[width=7cm]{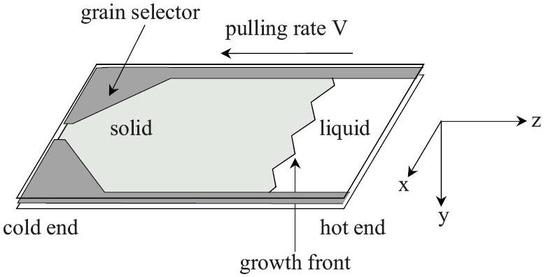}
\caption{Geometry of thin directional-solidification (T-DS) experiments.  
A 12 \mic-thick layer of liquid enclosed by glass plates and  polymer spacers is 
pulled toward the cold end in an imposed thermal gradient. The growth front is observed in real time  with a polarizing optical microscope. $z$: thermal-gradient  and growth axis. $y$ : normal to sample plane and direction of observation.  $x$: overall direction of the growth front.}
\label{setupf}
\end{figure}

This article reports the results of an experimental investigation of  pattern 
formation during  thin directional solidification (T-DS) in a substance 
forming faceted (monoclinic) crystals, namely, impure biphenyl. The geometry of the experiments is specified in Figure  \ref{setupf}.  We focus on the growth patterns,  called  weakly faceted (WF) patterns, that bring into play  only high-mobility facets. Under usual experimental conditions ({\it{i.e.}} far from any roughening transition), facet mobility is controlled by the motion of one-molecule-thick growth sk.jpg emitted from certain sites  (intersections with crystal dislocations, contacts with other crystals) \cite{Burton51}. A given facet can have a high, or a low mobility depending on whether, or not, it contains such step sources. The  high-to-low-mobility transitions of a facet during growth, if any,  are due to step sources entering or leaving the facet, and are quite sharp, and thus easily identified on a macroscopic scale  \cite{Chernov89}.

Interest in the  theory of  thin weakly faceted (WF) growth was aroused by experiments by Maurer {\it{et al.}} showing that  the facet length of free-growth dendrites of  NH$_4$Br followed the same $V^{-1/2}$  scaling law, where $V$ is the dendrite tip growth rate,  as the tip radius of nonfaceted  dendrites  \cite{Maurer89}. 
Ben Amar and Pomeau explained this finding by establishing analytically that the whole 
morphology of steady free-growth dendrites obeys a  $V^{-1/2}$ scaling law \cite{BenAmar88}
using a purely capillary  ($\beta=0$) 2D model of weak faceting  at low undercooling
$\Delta T$ of the liquid. 
These authors then showed that introducing standard facet kinetics ($\delta T_k\propto V^n$, where $n\le2$) in the model should not alter these conclusions. Adda Bedia and Hakim \cite{AddaBedia94,AddaBedia95} gave approximate analytical solutions for  free-growth dendrites with capillary facets.  Recently, Debierre and coworkers revisited this problem using phase-field numerical simulations \cite{Debierre03,Guerin05}, and extended the validity of  the  $V^{-1/2}$ scaling law  to arbitrary  $\Delta T$ and  capillary-anisotropy coefficients.  Concerning directional solidification, the main theoretical contributions so far are two analytical studies, one dealing with  the cellular transition in the particular case when a high-mobility facet is perpendicular to the growth direction by Bowley  {\it{et al.}}   \cite{Bowley89} and Caroli   {\it{et al.}}   \cite{Caroli89}, and the other dealing with  steady WF patterns at  $V>V_c$ in  crystals with two facets at $\pm45$\degres~from the growth axis by Adda Bedia and Ben Amar  \cite{AddaBedia91}. Among the numerous problems left unsolved today, most authors singled out the question of the respective roles of capillary and kinetic anisotropies in  WF growth. To tackle this problem is a current challenge for phase-field numerical simulations  \cite{Debierre03,Uehara03,boak2003}. 

With regard to experimental investigations capable of casting light on the dynamics of thin WF patterns, we are aware of studies dealing with mesophase  systems  \cite{meos90,bofa02, boak02}, but none of dealing with "solid" crystals. A reason for this is  the frequent occurrence of large-amplitude plastic deformations --called  growth-induced deformation (GID) thereafter-- breaking up faceted solid crystals into a multitude of small grains during T-DS. The origin of  GID, and the methods of keeping it from happening, if any, are still unclear. Fabietti and Trivedi  \cite{Fabietti97} studied GID during T-DS in impure naphthalene, which has  the same crystallographic structure as biphenyl, but their observations were inconclusive as regards the possible existence of deformation-free thin cellular patterns.
In this article, we  report that GID did not occur during thin free growth (T-FG) of biphenyl crystals indicating that GID basically is a thermal-stress effect generated by the externally applied thermal gradient. Most significantly,  we found that GID was also lacking during T-DS  when the (deformation-free) growth front was cellular. By triggering the cellular instability at an early stage of T-DS, it was thus possible to grow deformation-free steady  WF patterns  in impure biphenyl. We present a detailed study of the spatiotemporal dynamics of GID in  T-DS samples with single-crystal seeds. This study reveals that the first stage of GID is a long-range  process, which occurs only in large-width crystals, and not in the narrow cells of the cellular patterns. Finally, we report a first investigation of  thin WF patterns in deformation-free biphenyl crystals for various orientations of the crystal with respect to the growth direction.

\section{Materials and methods}
\label{exp}

\begin{figure}[htbp]
\includegraphics[width=7cm]{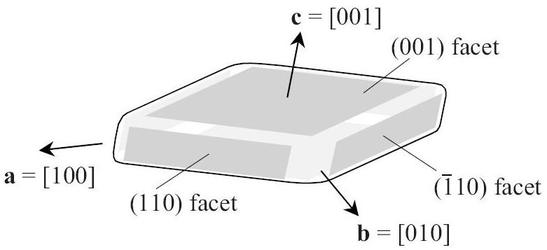}
\caption{Melt growth shape of a biphenyl crystal (schematic).}
\label{biphcryst}
\end{figure}

Biphenyl ($\rm{C}_{12}\rm{H}_{10}$) is a transparent substance, which crystallizes into  a biaxial birefringent monoclinic phase at $T_m\approx70$\degres C.  The point group of the crystal has a twofold axis   $\bf{b}$ and a mirror plane normal to  $\bf{b}$ that contains the other two lattice translations $\bf{a}$~and $\bf{c}$ (Fig. \ref{biphcryst}). The crystalline parameters  are  $a=0.81$, $b=0.56$ and $c=0.95$~nm, and the angle between  $\bf{a}$ and  $\bf{c}$  is 95.1\degres~ at room temperature \cite{LB1, LB2}.  Biphenyl crystals have a perfect (001) cleavage, and various glide systems involving dislocations with Burgers vectors  [100] and [010], but not [001] \cite{Marwaha92}. We show below that, during melt growth, biphenyl exhibits only   \{001\} ("basal"), and  \{110\}  facets. (It does not  exhibit \{100\}  facets, contrary to naphthalene).  The  [100] apex angle of the basal facet, {\it{i.e.}} the angle between  [110] and [1$\bar{1}$0], calculated from the above data is  $69.5$\degres. The terminology for the orientation of the crystals  employed hereafter is as follows. Crystals with (001)  parallel to the sample plane  are called (001)-oriented crystals. Their in-plane orientation is either specified by the angle $\theta_a$  of   $\bf{a}$ and  the growth axis (called $z$, see Fig. \ref{setupf}), or designated as  (001)[100]- or (001)[010]-orientation when $\theta_a=0$\degres~or $\theta_a=90$\degres, respectively. Crystals deviating from the (001)-orientation are called "tilted", and the angle of their basal plane and the sample plane is denoted $\omega$. Crystals with their basal plane nearly perpendicular to the sample plane are called "edge-on".

The observations were performed in commercial biphenyl (Fluka, 99.9 \%) unless otherwise mentioned. A few experiments were made with  biphenyl doped with 1 wgt\% of camphor. The methods of preparation and observation of the samples are explained elsewhere \cite{Mergy93,Akam96,Akamatsu95}, and need only to be briefly outlined here.  The crucibles consist of two  parallel glass plates separated by polymer spacers. Their inner dimensions are of 60mm along $z$,  8mm~(width) along $x$, and  12\mic~(thickness) along $y$. A funnel-shaped grain selector is created  near the cold end of the crucibles by using spacers of an  appropriate shape (Fig.  \ref{setupf}). The crucibles are filled  with the liquid compound under controlled Argon atmosphere,  quenched to room temperature, and sealed. They are then inserted into a  T-FG or a T-DS setup, and observed in side view  by videomicroscopy. Polarized light was used in order to take benefit of the birefringence of the crystals.

T-FG was  carried out  with a commercial hot stage (INSTEC, HS1-i) mostly for the purpose of preparing single-crystal seeds for T-DS. The temperature distribution in the setup is not perfectly uniform, but has a shallow depression at the center of the window of observation. This feature can be utilized to grow single-crystal seeds. The temperature of a polycrystalline sample is increased until all the grains melt except the one (the future seed) that is located at the minimum of the temperature distribution. The melting is pursued until the seed detaches itself from the glass plates, is carried away  by flows existing in the liquid, and redeposited with, generally, its basal facet closely aligned with the sample plane.
The temperature is then  decreased step by step in order  to make this (001)-oriented single crystal grow without (or with as little as possible) morphological instability. It should be noted that, during this process, the crystal is a thin plate limited by two blocked (deprived of step sources) basal facets, which are basically not  in contact with the glass walls. Some crystals had a small residual misalignment and collided with a glass wall during growth.  Crystals that filled the sample without hitting the glass walls  ($\omega<0.01$\degres)~ were selected as seeds for T-DS.  These crystals kept a uniform optical contrast during growth indicating that no noticeable GID occurred during T-FG. The fully solidified samples were then slowly cooled to room temperature, and transferred to the T-DS setup.  A weakly contrasted rectilinear striation parallel to the direction [010] appeared at  $\Delta T\approx$10K during cooling, and persisted in the non-melted part of the samples after they were inserted into the T-DS setup.  However, they were  not transmitted to the grown crystal during  T-DS 
indicating that they did not pertain to the bulk of the crystal, but to the layer of matter comprised between the crystal and the glass walls that solidified during cooling to room temperature. The crystallographic orientation of the striae and the presence of microbubbles are suggestive of slip bands  left by [010] dislocations gliding in the (100) plane. 

T-DS experiments were performed using a home-built stage made of two independent thermally regulated copper blocks separated by a several millimeters-wide gap, in which the solidification front sits. During this study,   $G\approx5.6~\rm{Kmm^{-1}}$, unless otherwise mentioned.  The translation velocity of the samples is stable to within $\pm 2 \%$ in the explored range (0.1-30\mis).  In addition to a standard T-DS stage, we used a new "rotating T-DS stage", to be presented in a future publication, with which it is possible to rotate the sample about  $y$, that is, to change the in-plane orientation of the crystal during solidification. The microstructure of the non-melted part (seed) of the sample at the beginning of  a T-DS run is an all-important experimental factor that can be largely controlled by an appropriate design of the  first stages of T-DS. The main alternative is to include, or not, a low-$V$ growth through the grain selector prior to the T-DS run proper. A few  as-quenched samples were solidified without grain selection, and exhibited a strongly faceted mode of growth.
As a general rule, as-quenched samples were solidified with grain selection, which led to a complete elimination of all the tilted grains (see Section \ref{GIPDasq}). 

A cellular instability was observed during T-DS of (001)-oriented grains under conditions, which will be specified later on. We measured $V_c$ at $G=5.6~\rm{Kmm^{-1}}$ in these grains by the method explained in Ref.  \cite{Mergy93}. Three types of samples must be distinguished: fresh  samples of commercial biphenyl, samples of commercial biphenyl having undergone a T-FG or T-DS solidification/melting cycle at low $V$ (as is the case of samples with  single-crystal seeds) designated as "purified" hereafter, and samples of camphor-doped biphenyl.  We found $V_c=4.5\pm1.5$\mis~in fresh undoped samples,  $V_c>15$\mis~in purified samples, and $V_c<1$\mis~in  camphor-doped samples. 

\section{Experimental results}\label{results}
\subsection{Growth facets of biphenyl crystals}\label{Facet}

\begin{figure}
\includegraphics[width=7cm]{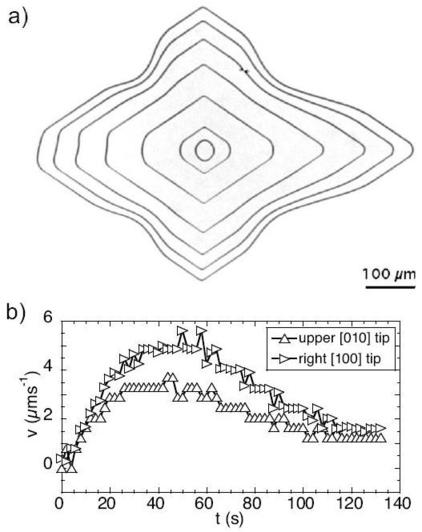}
\caption{a) ST diagram of  a (001)-oriented biphenyl crystal during T-FG. Time interval: 20~s. A quench  to $\Delta T\approx0.06$~K was applied to a quasi-circular seed at $t=0$. b) Growth rates of the [100] (rightmost) and [010] (uppermost) tips as a function of time.  The thermal time lag of the T-FG stage is about 1~min. The growth of the leftmost tip was perturbed by a dust particle.}
\label{biphfreeSeq}
\end{figure}

In T-FG as well as T-DS, biphenyl crystals exhibited only two types of facets, namely, low-mobility \{001\}  facets and high-mobility \{110\} facets. We give experimental evidence of the respective kinetics of these facets. Figure  \ref{biphfreeSeq}a shows a spatiotemporal (ST) diagram --{\it{i.e.}} a time series of binarized profiles  of the growth front displayed in the reference frame of the sample-- of  a  (001)-oriented biphenyl crystal  during T-FG. The crystal first grows from  a circular to  a rhombus shape limited by \{110\}  facets, then undergoes an impurity-driven (Mullins-Sekerka) instability, and finally settles into a steady dendritic regime with dendrites pointing in the $<100>$ and $<010>$ directions.  The  \{110\}  facets are linked to each other by smooth rounded regions indicating that no forbidden orientation range exists in the non-faceted parts of the solid-liquid interface. The faceted parts of the profile remained rectilinear within the measurement  uncertainty during the process.  The measured angle between [110] and [1$\bar{1}$0] facets was  of $67.5\pm0.6$\degres.  The small difference between this angle, and the one deduced from crystallographic data, if significant, is attributable to differences in temperature and composition. The slight crystallographic tilt of the \{110\} facets with respect to $y$ was not resolved, but a difference in contrast between the faceted and rounded part of the interface was apparent, revealing a difference (planar versus rounded) in 3d shape between these two regions.  The time evolution of the tip velocity $V$ (Figure  \ref{biphfreeSeq}.b) shows that  \{110\}  facets developed through a perfectly smooth process  indicating that no discontinuity in the kinetic coefficient is associated with   \{110\}  facets. The same  features  were also observed for   \{110\} facets during T-DS, as will be seen in Section \ref{stabil}.

\begin{figure}[htbp]
\includegraphics[width=8cm]{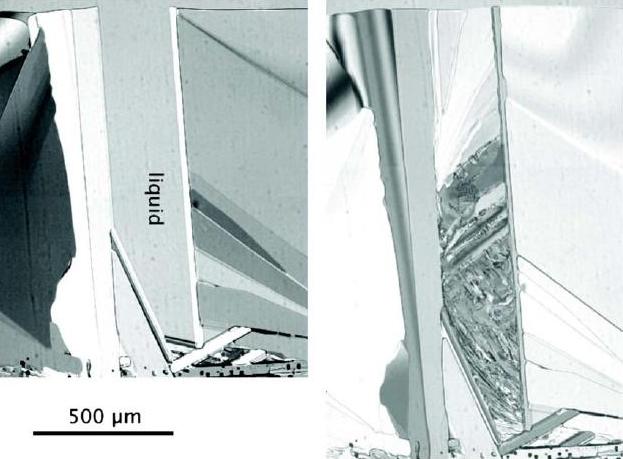}
\caption{T-DS of an as-quenched biphenyl sample consisting of many grains with different orientations at V = 6.5\mis. The growth direction is oriented upwards. Left:  liquid pocket at the rear of the growth front. Right: same area (with differently oriented polars) after the pocket was solidified by explosive nucleation.}
\label{nucleation}
\end{figure}

Detailed information about the growth kinetics of  \{001\}  facets was given by T-DS experiments performed without grain selection in as-quenched samples. Figure \ref{nucleation} shows a deep liquid pocket due to the mutual impingement of edge-on crystals adhering to the glass walls through their nonfaceted extremities. A sporadic nucleation of macrosk.jpg on the basal facets  bordering the liquid pocket took place as $\Delta T$ progressively increased until the whole pocket was suddenly filled by a polycrystal through an "explosive" nucleation process.  The distance of the nucleation site from the growth front  was of about~1.5mm corresponding to  a temperature difference of about 9K. 

In conclusion, the known absence of dislocations capable of serving as step sources for (001) facets --namely, dislocations with a  [001] Burgers vector-- in biphenyl crystals  is a sufficient explanation of the low mobility of these facets. Reciprocally, the observed high mobility of  \{110\}  facets must be attributed to  dislocations with [100] or [010] Burgers vector intersecting these facets. The only acceptable alternative would be the proximity of a roughening transition, but this seems very unlikely given the large extension of these facets, and the small growth rates considered in this study. However, the mechanisms by which   \{110\}  facets are fed with dislocations during growth even when the growth morphology becomes very complex, are unknown. Anticipating on the observations presented below, we note that one, and probably the most important of these mechanisms is the plastic deformation generated by thermal stresses illustrated by GID. The formation of a stratified microstructure, or equivalently of (001) twist boundaries, during growth explained in Section \ref{GIPDsingle} may also contribute, but  it is not certain that these boundaries survive the morphological transitions of the system (see Fig. \ref{feuillets} below).

\subsection{Growth-induced plastic deformation during T-DS}
\label{GIPD}

\subsubsection{GID in samples with  single-crystal seeds}
\label{GIPDsingle}

We performed T-DS experiments in a series of samples with single-crystal seeds with various in-plane orientations at values of $V$ ranging from 0.5 to 30\mis. In these samples, $V_c\approx15$\mis, as previously mentioned.   We observed GID processes with similar characteristics in all the experiments. This, and the fact that GID does not occur during T-FG, may suffice to establish that GID is not due to a cellular instability, but most probably to thermal stresses. However, to substantiate this conclusion, and because of the remarkable features of GID in our experiments, we study the geometry of this process in detail.
 
\begin{figure}[ht]
\includegraphics[width=9cm]{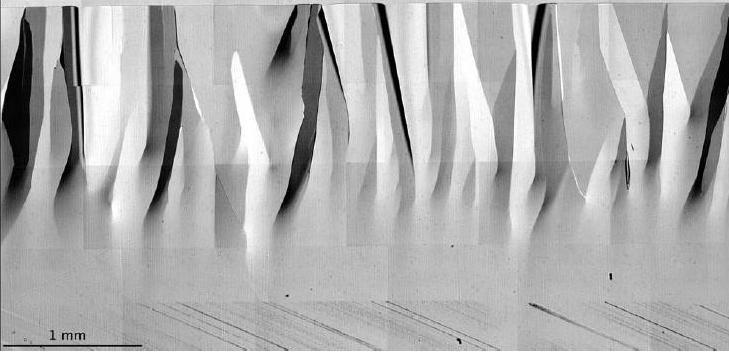}
\caption{GID during T-DS from a single-crystal  seed with $\theta_a=24$\degres.  $V=6.5$\mis. The seed can be recognized from the [010] striation inherited from the post-T-FG cooling down of the sample.}
\label{GID24}
\end{figure}

\begin{figure}[ht]

\includegraphics[width=8cm]{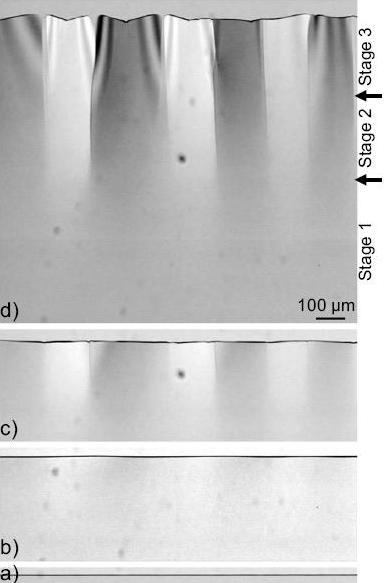}
\caption{GID during T-DS from a nearly (001)[100]-oriented single-crystal seed ($\theta_a=-2.5$\degres).  $V=30$\mis. {\bf a}) t=0~s (start of T-DS). {\bf b}) t=22~s. {\bf c}) t=33~s. {\bf d}) t=46~s.   
The two arrows indicate the appearance of grain boundaries and the first collisions with the glass plates, respectively.}
\label{GIDA}
\end{figure}

\begin{figure}[ht]
\includegraphics[width=8cm]{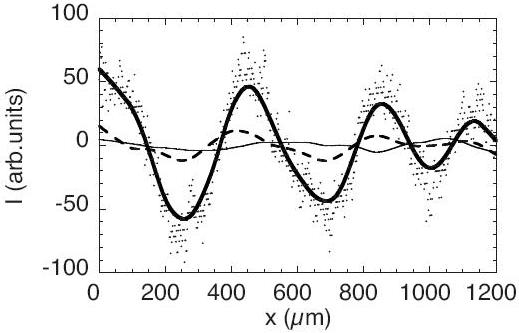}
\caption{Gray-level plots measured at different solidification length $L_s$ during the T-DS run 
shown in  Fig. \ref{GIDA}.  Thin line:   $L_s=0$. Broken line:  $L_s=100$\mic. 
Thick line and dots:  $L_s=200$\mic. The curves were obtained by smoothing the data points
(dots) after subtraction of  a linear function fitted onto the background.}  
\label{levelPlots}
\end{figure}

GID processes that took place in  two different crystals oriented, one asymmetrically, and the other symmetrically, with respect to the T-DS setup are shown in Figures \ref{GID24} and  \ref{GIDA}, respectively. Both processes go through  the following three successive stages: Stage 1, which starts from the beginning of the solidification, and consists of the progressive amplification of a smooth contrast modulation extending along $x$ (Fig. \ref{levelPlots});  Stage 2, which begins with the sudden creation of grain boundaries  at positions corresponding to the extrema of  the contrast modulation; Stage 3, during which crystal-glass collisions followed by the appearance of new grains lead to a fully polycrystalline microstructure.

\begin{figure}[ht]
\includegraphics[width=5cm]{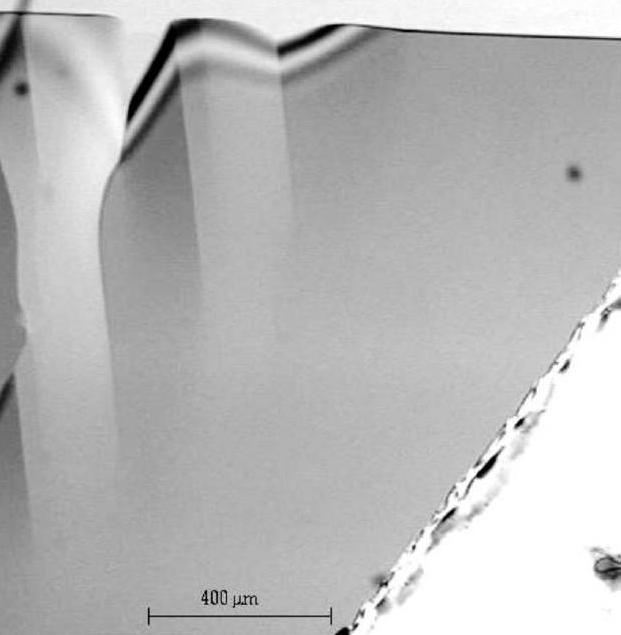}
\caption{GID during T-DS from an as-quenched seed with $\theta_a=-6^o$. $V$=1.55\mis  
. $\Lambda_x = 220\mu m\pm20\mu m$. 
The slanting edge of a grain selector appears in the lower-right hand corner of the photograph.}
\label{funnel}
\end{figure}

\begin{figure}[ht]
\includegraphics[width=7cm]{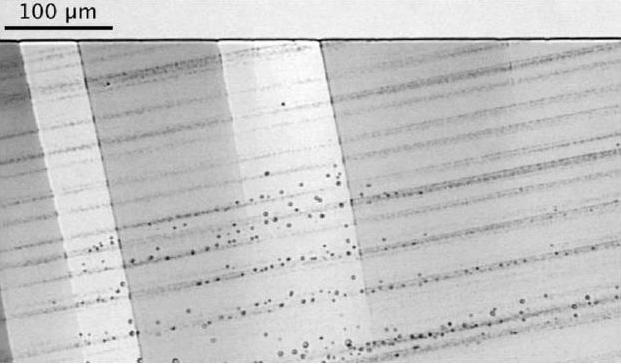}
\caption{GID during T-DS from a single-crystal seed with $\theta_a=-10$\degres. $V$=0.3\mis. After T-DS arrest, the heating of the ovens was switched off for a few hours, and then put on again.  
The grain boundaries are perpendicular to the [010] striation and thus parallel to [100]. 
}
\label{GIDstria}
\end{figure}

\begin{figure}[ht]
\includegraphics[width=8cm]{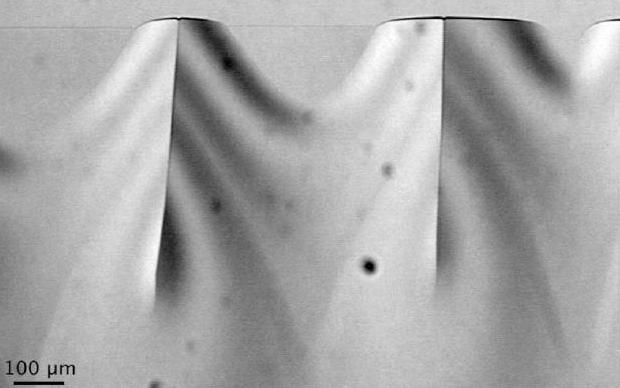}
\caption{GID during a second T-DS run in the same sample as in Fig. \ref{GIDA}.  
$V=0.54$\mis . $\Lambda_x = 540\mu m\pm20\mu m$. 
}
\label{GIDB}
\end{figure}

Stage 1 clearly is a long-range process of plastic deformation of the growing 
crystal. A full understanding of such a process under the conditions of our experiments
(strong confinement of the system and elasto-plastic anisotropy of the solid) is 
notoriously beyond reach at present, but some interesting semi-quantitative remarks 
can be made. Various observations indicated that the distortion field existing in the 
crystal during Stage 1 mostly consisted of rotations of the crystal lattice about the 
[100] axis of the crystal. Such a distortion field favors the formation of tilt 
boundaries (arrays of [010] edge dislocations) in (010) planes, in agreement with the 
fact that the grain boundaries appearing at the onset of Stage 2  were approximately 
parallel to [100] in all the samples studied (Figs. \ref{funnel} and \ref{GIDstria}). 
Figs. \ref{GIDA} and \ref{levelPlots}, on the one hand, and Fig. \ref{GIDB}, on the 
other hand, correspond to two successive runs 
performed in the same (001)[100]-oriented sample at $V=30$\mis and $0.54$\mis, 
respectively. The crystal length solidified during the first run was entirely re-melted 
before the second run so that there was no possible influence of the first GID on the 
second one. We note that  $V$ was above the cellular instability threshold during the 
first run, which is unimportant for our present purpose, but has interesting 
consequences, which will be commented later on. During these experiments, Stage 1 clearly
exhibited two characteristic lengths, namely, a wavelength $\Lambda_x$ along $x$, and a 
solidification length $L_z$ along $z$. During both runs, $\Lambda_x$, defined as the 
spacing of the extrema of grey-level plots (Fig. \ref{levelPlots}), ranged from $300$ to 
$550$\mic. 
This scatter was mostly due to the existence of a lateral gradient of unknown origin  (but 
probably linked to some experimental imperfection). The value of $L_z$, defined as the solidification length
between the start of the experiment and the first appearance of grain boundaries, was of 
$500\pm 50$\mic \ in both experiments. Larger ranges of $\Lambda_x$ and $L_z$ were found 
in asymmetrically oriented crystals (Figs. \ref{funnel} and \ref{GIDstria}) than in the 
(001)[100]-oriented ones, 
but the orders of magnitude remained the same. These observations indicate that 
$\Lambda_x$ and $L_z$ were essentially independent of $V$, as could be expected from the
fact that the plastic deformation started from the beginning of the solidification. 
This suggests that these lengths are mostly determined by the geometry of the 
experiment.

\begin{figure}[ht]
\includegraphics[width=7.5cm]{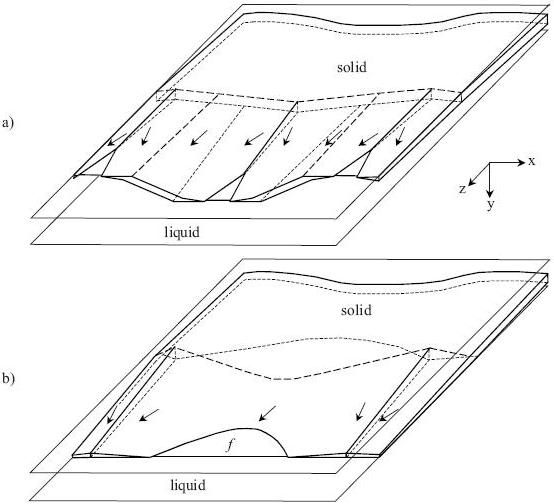}
\caption{Schematic 3D reconstruction of the crystal microstructures of Figs. \ref{GIDA} ({\bf a}) and \ref{GIDB} ({\bf b}). {\it{f}}: thin film crystal. Continuous lines: interfaces and grain boundaries. Broken lines:  zones of rapid, but continuous change in orientation. Arrows:  local orientations of the axis [100].  The sketches show a single period (along {\bf x}) of the GID microstructures.}
\label{croquis}
\end{figure}

Contrary to Stage 1,  Stages 2 and 3 presented features, which depended on the 
in-plane orientation of the crystal, and on additional, ill-known, geometrical 
factors. This point is illustrated in Figs. \ref{GIDA} and \ref{GIDB}, which show 
GID processes that occurred  during two successive runs  at different values of 
$V$ in the same (001)[100]-oriented sample. The common features of the two runs 
are that  the GID-induced microstructure kept the mirror symmetry about  $yz$ of the initial crystal  as well as the periodicity inherited from Stage 1 (in contrast with what occurred in asymmetrically oriented samples), and that thickness fringes appeared  near the grain boundaries at, or a short time after, the onset of Stage 2. The most apparent differences between the two runs are the additional symmetry (twofold axis at $\Lambda_x/4$) existing in Fig. \ref{GIDA} compared to Fig. \ref{GIDB}, and the fact that the thickness fringes that appear on either side of a grain boundary are divergent  in Fig. \ref{GIDA}, but convergent in Fig. \ref{GIDB}.  Thickness fringes can only arise from ultra-thin crystal wedges. Figure \ref{croquis} displays a schematic 3d reconstruction of the  microstructures based  on this remark, and the following additional conjectures:  (i) the onset of Stage 2 occurred when the ongoing deformation made the crystal come into contact with the glass plates. At each point of contact, a pair of misoriented crystal wedges attached to the glass were created under the effect of external forces (linked with capillarity, flows in the liquid, changes in the thermal field); (ii)  the initial positioning of the growing crystal with respect to the glass plates  was symmetric in  Fig. \ref{GIDA}, but strongly asymmetric in  Fig. \ref{GIDB}. Regardless of the details, Fig. \ref{croquis} illustrates the difference in nature between the early stages of GID,  which do not involve contacts or collisions with the container walls, and later stages, which are essentially driven by such events. The specificity of GID, in the sense given to this term here,  lies in Stage 1, while collision-induced deformations similar to those occurring during Stage 3 are also generated by a misaligned seed.

The nature of the growth process  following crystal-glass collisions is most clearly  illustrated in  Figs. \ref{GIDB} and \ref{funnel}. Without discontinuity, and thus, probably through a  plastic deformation of the crystal colliding the glass, solidification continues with the growth of (001)-oriented thin-film crystals in contact with, or very close to the glass walls. The extreme thinness of these crystals is revealed by a slight recoil of their solid-liquid interfaces with respect to thicker parts of the crystal attributable to their strong curvature in the direction $y$ (Gibbs-Thomson effect). 

\subsubsection{Grain growth and GID in as-quenched samples}
\label{GIPDasq}

\begin{figure}[htbp]
\includegraphics[width=7.5cm]{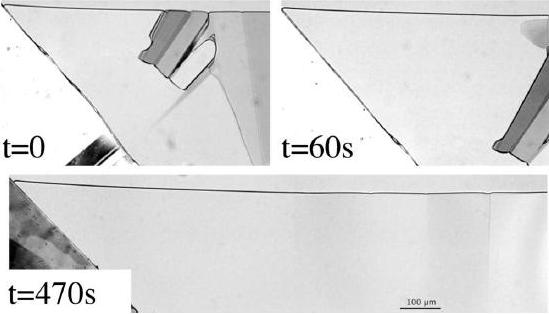}
\caption{Grain growth and onset of GID  during the early stages of T-DS from an as-quenched seed.  $V = 3.1~ \mu \rm{ms^{-1}}$.  The slanting edge of a grain selector is visible in the lower-left hand corner of the photographs.}
\label{rearr}
\end{figure}

\begin{figure}[htbp]
\includegraphics[width=6cm]{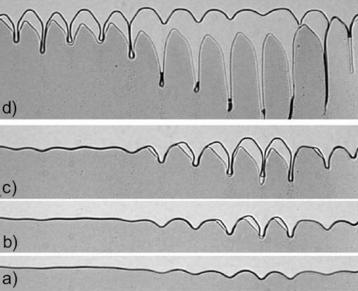}
\caption{Disclosure and elimination of a stratified microstructure during a cellular-instability transient.  $V= 4.7$\mis.  a) $t=0$; b) $t=34$~s; c) $t=40$~s and d) $t=56$~s. The difference in orientation of the stratums is about 8\degres. Horizontal dimension: $620 \mu m$.}
\label{feuillets}
\end{figure}

A spontaneous grain-growth process leading to a complete elimination of tilted grains in favor of (001)-oriented grains took place in as-quenched samples  during the grain-selection stage of T-DS  (Fig. \ref{rearr}). When their size permitted it, the (001)-oriented grains emerging from this process underwent a GID process similar to those observed in samples with single-crystal seeds.  The GID process repeated itself with a constant amplification rate during the continuation of the growth through the funnel-shaped selector (Fig. \ref{funnel}). Adjacent (001)-oriented grains competed and often overlapped,
leading to  stratified microstructures (Fig. \ref{feuillets}).

\subsubsection{Cellular instability and GID}
\label{GIPcell}

The rapid development of GID prevented us from observing deformation-free WF patterns in samples with  single-crystal seeds. To bring about cellular transitions in deformation-free crystals, we applied upward $V$-jumps to  (001)-oriented grains at an early stage of GID  in fresh undoped samples. Quite significantly, we never observed GID to appear after such a cellular transition occurred.  In camphor-doped samples, cellular transitions without GID  were observed at all the explored values of $V$.

Cellular-instability transients occurring concurrently with GID processes are worth briefly considering as a new instance  of coupling between plastic deformation in the solid and impurity-driven dynamics at the growth front \cite{Bottin02}.  It is known that impurities rejected by the growing edge of a plate-like crystal partly segregate at the rear of this edge \cite{Fujioka74}. In the confined geometry of T-DS, this effect manifests itself by an  increase of the equilibrium temperature, and thus an advance of the growth front, inversely related to the crystal thickness. Such a relation between the profile $z=\zeta(x)$ of the growth front and the local value of the crystal thickness was observed during GID except in the thinnest regions of the front where  the Gibbs-Thomson effect predominated. It should be noted that this  impurity-driven effect, which remains small at low $V$, amplifies as  $V$ approaches $V_c$ (Fig. \ref{GIDA}) conferring an imperfect-bifurcation character to the cellular instability.

\subsubsection{Origin of GID}
\label{origin}

In conclusion, the core of the GID process is the  progressive amplification of a long-range modulated plastic deformation of the growing crystals (Stage 1). The subsequent stages of the process are essentially geometrical consequences of this first stage. The lack of sensitivity of Stage 1 to the control parameters of the solidification rules out the possibility that this process be strongly connected with an  impurity-driven dynamics. The fact, that in our system the growing crystals that underwent GID were not in contact with the container walls makes it very likely that the first stage of GID  is basically due to the thermal gradient alone --more precisely, the stresses engendered by an inhomogeneous temperature field in a crystal with strong  elastic and plastic anisotropies.  This stress field  depends only on the geometry of the growing crystal at fixed geometry of the T-DS setup. This is consistent with the observed insensitivity of $\Lambda_x$ and $L_z$ to  $V$ and $\theta_a$, and suggests that the lack of GID in cellular fronts is simply due to the fact that stresses and plastic deformation are not transmitted from a cell to another. 

\subsection{Weakly faceted cells and dendrites} \label{WFPs}

\subsubsection{Stability of weakly faceted patterns} \label{stabil}

\begin{figure}[ht]
\includegraphics[width=6cm]{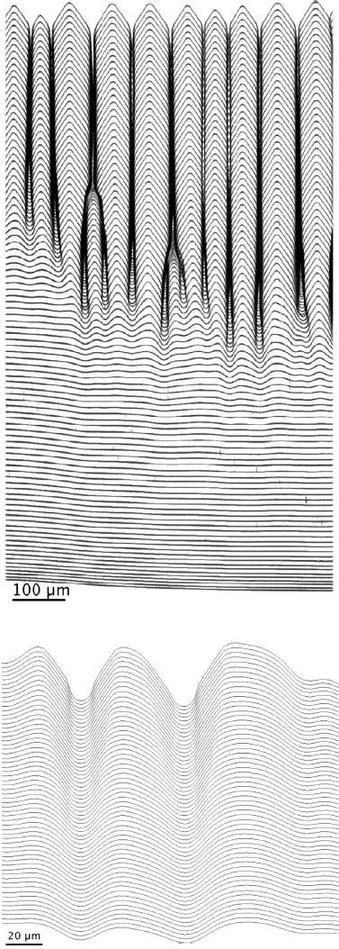}
\caption{ST diagram  of  a cellular-instability transient during T-DS of a (001)[100]-oriented grain. Time interval: 2~s (top), 0.08~s (bottom).  $V$=6.5~\mis.}
\label{diffinstab}
\end{figure}

\begin{figure}
\includegraphics[width=6cm,height=5.9cm]{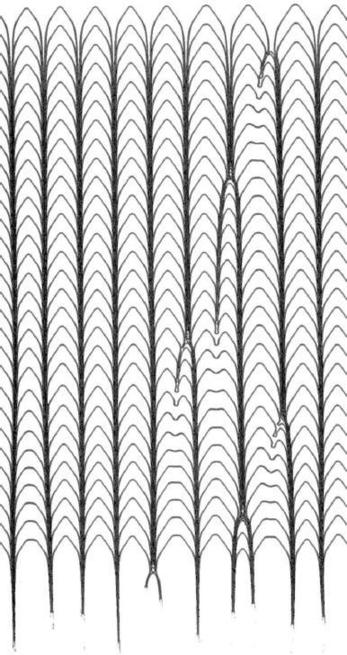}
\caption{ST diagram (time intervals 2~s) of a propagating  tip-splitting instability during T-DS of  a (001)[100]-oriented grain. $V$=17\mis.  Note the transient oscillations in the wake of the solitary wave. Horizontal dimension: $620~\mu m$. The diagram has been contracted by a factor of 2 vertically. }
\label{splitting}
\end{figure}

Like their non-faceted homologs, the WF patterns of biphenyl exhibited  a broad  ($40-100$\mic, typically) range of stable spacings $\lambda$  at fixed $V$. This range was bordered by a cell elimination instability at small $\lambda$  and a cell splitting instability at large $\lambda$. We observed cell elimination processes during cellular-instability transients  (Fig. \ref{diffinstab}), and could estimate the  cell elimination threshold spacing to be about 40\mic~in undoped biphenyl in the explored $V$ range ($V/V_c<4$).  On the other hand, we noted various modes of instability at  $\lambda>100$\mic, in particular, oscillations and a propagating tip splitting  (Fig. \ref{splitting}). 

\begin{figure}[ht]
\includegraphics[width=7cm]{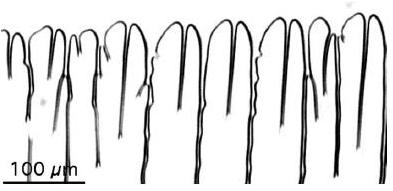}
\caption{{\it{Doublons}} during T-DS of a  (001)-oriented grain. Camphor-doped biphenyl. $V$=150\mis.}
\label{biphdoublons}
\end{figure}

In their recent numerical study of the free growth of  thin WF systems in a channel, Gu\'erin   {\it{et al.}}  identified an oscillatory symmetry-broken mode of growth, and argued that this mode belongs to the same branch of states as  the  steady asymmetric fingers that exist in low-anisotropy nonfaceted systems \cite{Guerin05}. In T-DS, such fingers, if any, should appear in the form of pairs of  asymmetric fingers called {\it{doublons}}  \cite{Ihle94,Akamatsu95,Utter01}. We performed a few experiments at very high  $V/V_c$  in camphor-doped biphenyl samples, and indeed observed a dendrite-to-{\it{doublon}} transition    (Fig. \ref{biphdoublons}) lending support to Gu\'erin   {\it{et al.}}'s argument. It must be noted, however, that we have not been able to ascertain the existence of facets at the tips of theses  {\it{doublons}}, so that the possibility of a roughening transition occurring in our system at high $V$ cannot be entirely excluded.

\subsubsection{(001)[100]-oriented grains in undoped biphenyl} \label{symm}

\begin{figure}[ht]
\includegraphics[width=7cm]{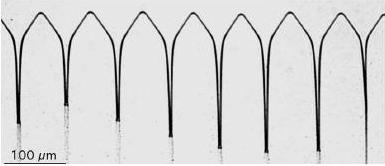}
\caption{Cellular pattern during T-DS of a [001](100)-oriented crystals. Undoped biphenyl. $\theta_a\approx1$\degres. $V$ = 6.5 \mis. }
\label{100cells}
\end{figure}

\begin{figure}[htbp]
\includegraphics[width=7cm]{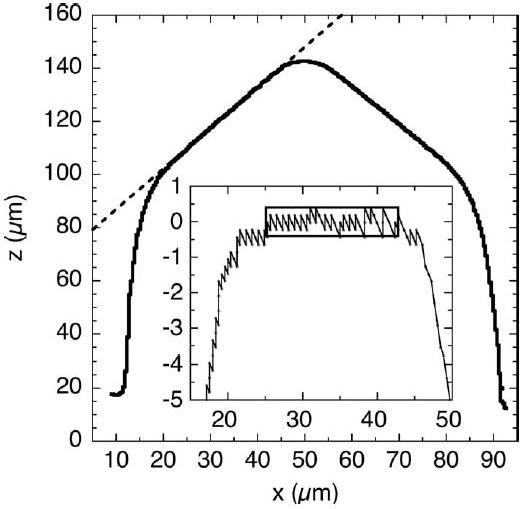}
\caption{Determination of the length and orientation of a facet in a T-DS cell. The profile was extracted from Fig. \ref{100cells}a by thresholding and skeletonizing the image of a cell.  The broken line is the linear best-fit function along a presumed facet. This function was subtracted from the relevant part of the profile yielding the set of data points displayed in the inset. The box encloses the data points whose deviation from a smooth curve drawn through these points is smaller than their scatter. The length of the box (corrected for the projection factor) was taken as a measure of the facet length.
}
\label{facet_def}
\end{figure}

\begin{figure}[htbp]
\includegraphics[width=6.9cm]{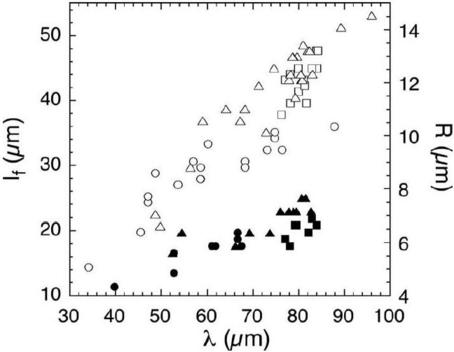}
\caption{Facet length $l_f$ (open symbols)  and tip radius $R$ (filled symbols) versus cell spacing $\lambda$  during T-DS of (001)[100]-oriented crystals. Undoped biphenyl. Squares: $V=$4.7\mis. Triangles:  $V=$6.2\mis. Disks: $V=$17\mis.}
\label{l_facet}
\end{figure}

Cellular patterns  were symmetric ({\it{i.e.}} did not  drift in the direction $x$) in (001)[100]-oriented grains of undoped biphenyl (Fig. \ref{100cells}), as could be expected since a mirror plane of the crystal structure is parallel to  $yz$  in these grains. The  $\lambda$-distribution ranged  from about 40 to 100\mic~at the end of the cellular transition, and was slowly relaxing toward a uniform distribution. It was not possible to wait for this relaxation to be complete because the lifetime of the grains of interest was limited by the competition of the adjacent grains. We performed measurements in quiet regions of the evolving patterns assuming the quasi-steady-state condition  to be valid in these regions.  We measured the cell facet length and tip radius during three different T-DS runs in  (001)[100]-oriented crystals of undoped biphenyl by the method explained in the legend to Figure \ref{facet_def}.  The results are displayed in Figure \ref{l_facet}. The measurement error on $l_f$ was of $\pm2$\mic, and thus of $10\%$, at  worst, which can account for the scatter of the data.
The major trend emerging from the data is an essentially linear increase of $l_f$ with
increasing  $\lambda$.
The concomitant increase in $R$, which is precursor to tip splitting, is much weaker. The dependence of  $l_f$ and  $R$  on  $V$ is undetectable. The fraction of the front occupied by facets ($\approx0.5$) is important, and increases as  $\lambda$ increases. The sidebranching threshold is thus shifted  to high values of  $\lambda$  compared to what it would be without facets. This effect is most clearly illustrated  by the long distance separating the first sidebranches from the tips of tilted dendrites  in Figure \ref{FacetPerp} below.

\subsubsection{WF cellular patterns as a function of in-plane orientation in camphor-doped biphenyl}

\begin{figure}[ht]
\includegraphics[width=7cm]{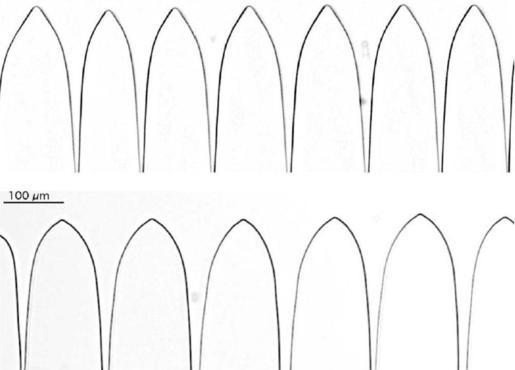}
\caption{Cellular  patterns during T-DS in camphor-doped biphenyl.  Above: (001)[100]-oriented crystal. $V =10.0$\mis; below: (001)[010]-oriented crystal. $V =5.0$\mis.}
\label{celldoped}
\end{figure}

\begin{figure}[ht]
\includegraphics[width=7cm]{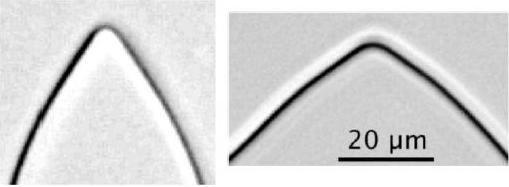}
\caption{Enlargement of cell tips from  Fig. \ref{celldoped}. Left: [100] cell. Right:  [010] cell.}
\label{highmag}
\end{figure}

\begin{figure}[htbp]
\centerline{\includegraphics[width=6.5cm]{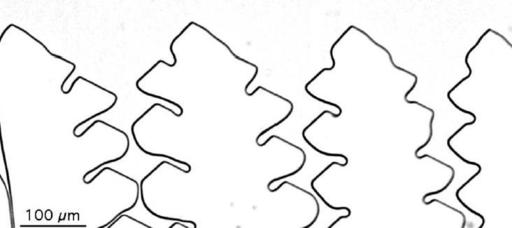}}
\caption{[010] dendrites during T-DS of a (001)-oriented grain with  $\theta_a$= 82\degres.  Camphor-doped biphenyl. $V$=10\mis. The  tilt angle of the dendrites is  $8\pm1$\degres.}
\label{tilted010dend}
\end{figure}

The grain-growth process studied in Section \ref{GIPDasq} yielded (001)-oriented grains with in-plane orientations belonging to a limited interval around $\theta_a=0$. We  used a  rotating T-DS stage to grow (001)-oriented crystals with   arbitrary $\theta_a$  values  in camphor-doped biphenyl samples. The uncertainty on  $\theta_a$ was of $\pm0.2$\degres. At all  $\theta_a$ values cell tips exhibited well-defined  facets, which were however too small to permit a quantitative study. We observed a lateral drift of the cellular patterns for all the in-plane orientation of the crystal except the (001)[100]- and  (001)[010]-orientations   (Fig.  \ref{celldoped} and \ref{highmag}). In the last-named orientation, the crystal structure is not invariant to reflection with respect to the $yz$ plane, but the absence of drift indicates that the whole system has such an invariance. (It should be noted that monoclinic angle of biphenyl is small). In other words, the behavior of the system was practically that of a 2D system with two orthogonal symmetry axis ([100] and [010]).  It is worth noting that, contrary to [100] cells, [010] cells did not exhibit tip splitting but transformed to dendrites at large spacing values (Fig.  \ref{tilted010dend}).

We studied the lateral drift of the cellular patterns in grains with $\theta_a\neq 0$  and $\theta_a\neq\pi/2$. Like in  nonfaceted systems \cite{Akamatsu97, GeorgelinXX},  the tilt angle of the direction of growth of the cell tips  was an increasing function of $V$ and approached the  tilt angle of the  nearest axis of symmetry of the system ($\theta_a$  or $\pi/2-\theta_a$)  at large values of $V/V_c$ (Fig. \ref{tilted010dend}).

\begin{figure}[ht]
\includegraphics[width=8cm]{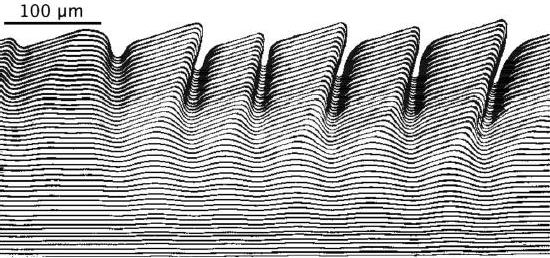}
\caption{ST diagram (length of sequence 48s) of a cellular-instability transient during T-DS of  a (001)-oriented grain with $\theta_a=40\pm1$\degres.  Undoped biphenyl. $V$=6.2 \mis. 
}
\label{drift_cells}
\end{figure}

\begin{figure}[htbp]
\includegraphics[width=7cm]{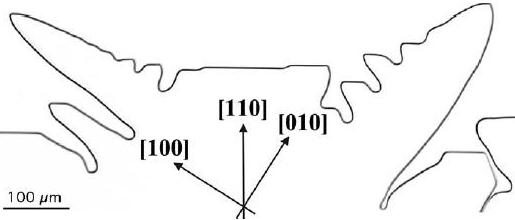}
\caption{Cellular-instability transient during T-DS of a (001)-oriented grain with a  \{110\} facet nearly perpendicular to $z$. Camphor-doped biphenyl. $V$=10\mis. The angle of the facet to $z$ is about 88.5\degres.}
\label{FacetPerp}
\end{figure}

Figure \ref{drift_cells} shows the ST diagram of a cellular-instability transient in a crystal with $\theta_a=40\pm1$\degres. Though the shape of the cells rapidly departed from a mere sine, no drift of the structure was observed until facets appeared. Given that kinetic anisotropy controls the drift velocity during the first stages of the cellular transient \cite{Coriell76}, this indicates that this anisotropy is relatively weak for {\bf n} belonging to the basal plane  of biphenyl. Finally, we note that grains with  a \{110\} facet nearly perpendicular to the growth axis exhibited a singular dynamics displaying a coexistence between crenellated interfaces \cite{Bowley89,Caroli89,meos90},  [100]-dendrites and [010]-dendrites  during cellular-instability transients (Fig. \ref{FacetPerp}). 

\section{Conclusion}
\label{conclu}

A thin plate-like crystal confined between two walls and submitted to  directional solidification is prone to  plastic deformation under the effect of thermal stresses generated by the applied thermal gradient, and  interactions with the walls. This study has shown that a deformation-free directional solidification of plate-like crystals is feasible, at least in the case of impure systems, and provides a good experimental model of the dynamics of 2D weakly faceted directional solidification. We have presented first elements of an experimental investigation of this dynamics, including a set of preliminary quantitative data about the facet lengths of  steady WF cellular patterns as a function of $\lambda$ and $V$. Previous experimental and theoretical studies have established that the scaling laws of nonfaceted dendritic free growth still hold in weakly faceted systems. Similarly, although in a less precise way, this study has shown that the dynamics of weakly faceted  directional-solidification systems is similar to that of  anisotropic, but nonfaceted systems, except, perhaps, when the growth front is nearly parallel to a facet. Theoretical studies  pointed out that the length of the facets near dendrite tips, which is directly connected to the capillary and kinetic coefficients of the solid-liquid interface,  is the most specific feature of WF growth morphologies. Further experimental and numerical studies of the   facet lengths of  steady WF cellular patterns as a function of various control, and material parameters could cast light upon basic questions such as the  respective roles of capillary and kinetic anisotropies in  weakly faceted growth.

\begin{acknowledgments}
T.B. ackowledges support by the European Community 
(Contract No. HPMF-CT-1999-00132) 
and by the Hungarian Scientific Research Fund (Contract No. OTKA-K-62588).

\end{acknowledgments}

\end{document}